\documentclass[prl,twocolumn,twoside,showpacs,superscriptaddress,floatfix]{revtex4}
\usepackage{graphics,amssymb,amsmath,epsfig,color}
\epsfclipon

\begin{document}
\title{Link-disorder fluctuation effects on synchronization in random networks}

\author{Hyunsuk Hong}
\affiliation{Department of Physics and Research Institute of Physics and Chemistry, Chonbuk National University, Jeonju 561-756, Korea}
\author{Jaegon Um}
\affiliation{School of Physics, Korea Institute for Advanced Study, Seoul 130-722, Korea}
\author{Hyunggyu Park}
\affiliation{School of Physics, Korea Institute for Advanced Study, Seoul 130-722, Korea}

\date{\today}

\begin{abstract}
We consider one typical system of oscillators coupled through disordered link
configurations in networks, i.e.,
a finite population of coupled phase oscillators with distributed intrinsic frequencies
on a random network.
We investigate collective synchronization behavior, paying particular attention to
link-disorder fluctuation effects on the synchronization transition and its finite-size
scaling (FSS). Extensive numerical simulations as well as the mean-field analysis
have been performed.
We find that link-disorder fluctuations effectively induce {\em uncorrelated random}
fluctuations in frequency, resulting in the FSS exponent $\bar\nu=5/2$, which
is identical to that in the globally coupled case (no link disorder) with frequency-disorder
fluctuations.
\end{abstract}
\maketitle

\section{I. Introduction}
Recently, it has been reported that disorder fluctuations play a crucial role on the
critical scaling near the onset of synchronization, even in the
mean-field (MF) regime~\cite{ref:Hong_Entrainment, ref:gl_noiseless, ref:Daido}.
In particular, one recent study on the collective synchronization in the system of globally
coupled oscillators (Kuramoto model) reports that ``sample-to-sample fluctuations''
induced by different realizations of intrinsic random frequencies yield significantly
different finite-size scaling (FSS) from the ordinary MF prediction~\cite{ref:Hong_Entrainment}. The anomalous fluctuations are characterized by the FSS exponent $\bar\nu=5/2$~\cite{ref:Hong_Entrainment}. When the random disorder in distributed intrinsic frequencies is completely removed, fluctuations are still anomalous but much weaker
($\bar\nu\simeq 5/4$)~\cite{ref:gl_noiseless}, because dynamic (temporal) fluctuations dominate in this case.
We call the latter as the {\em noiseless} frequency distribution, while the former as the {\em noisy}
frequency distribution.

When the system is put on a complex network, we need to consider another kind of fluctuations, i.e., ``link-disorder fluctuations'' induced by different configurations of connectivity among vertices in the network.
Considering that many systems in nature are based on a complex network topology,
it would be interesting to investigate how the link-disorder fluctuations affect the
collective behavior of a given system in general.  Most of collective behaviors invoking
phase transitions in networks belong to the conventional MF class where dynamic (thermal)
MF fluctuations are dominant, thus sample-to-sample disorder fluctuations become
usually irrelevant~\cite{ref:Hong_Entrainment, ref:Critical_phenomena_complex_networks_RMP, ref:sync_network_MF, ref:SFN_sync_Hong, ref:Ising_network_MF, ref:XY_network_MF, ref:phenomenology_critical_behavior_in_network, ref:Finite_size_scaling_network, ref:Finite_size_scaling_network_fractal_properties}.
However, in the presence of additional distributed quantities at vertices, link-disorder
fluctuations would generate {\em effective} disorder fluctuations in distributed quantities,
which might yield non-MF fluctuations.

In this study, we consider the system of coupled oscillators on a sparse link-disordered network. To separate out the effects of uncorrelated random link-disorder fluctuations, we focus on the system with a {\em noiseless} intrinsic frequency distribution on the Erd\"{o}s-R\'{e}nyi (ER) random and $z$-regular network~\cite{ref:Erdos}. The FSS property is investigated through the phase synchronization order parameter in extensive numerical simulations and
the value of the FSS exponent $\bar\nu$ is estimated. We find that $\bar\nu\simeq 5/2$, which is identical to that of
the globally coupled model with the {\em noisy} frequency distribution.
This implies that the random link-disorder fluctuations effectively generate random and independent
disorder in the frequency distribution. It is also found that the addition of random disorder in the intrinsic frequency distribution (noisy case) on the ER network does not change the
FSS exponent.

The present paper is outlined as follows.  In Sec.~II we introduce the model, and
explain how we generate the noiseless frequency distribution.
In Sec.~III, we present the mean-field (MF) analysis on the ER network with the predictions on the order parameter exponent $\beta$ and the FSS exponent $\bar\nu$.
Section~IV shows numerical simulation results.  A brief summary is given in Sec.~V.

\section{II. Model}
We begin with a finite population of $N$ coupled phase oscillators
on a sparse network.
To each vertex $i$ of the network, we associate an
oscillator whose state is described by the phase $\phi_i$ governed by
the equation of motion
\begin{equation}
\dot\phi_i = \omega_i - K\sum_{j=1}^N a_{ij} \sin(\phi_i - \phi_j),
\label{eq:model}
\end{equation}
where $\omega_i$ represents the intrinsic frequency of the $i$th oscillator.
It is assumed that $\{\omega_i\}$ are distributed according to a symmetric and unimodal
distribution function $g(\omega)$ such as a
Gaussian one with zero mean $(\langle \omega \rangle=0)$ and finite variance
$(\langle \omega_i \omega_j\rangle=\sigma^2 \delta_i \delta_j)$.
The $\{a_{ij}\}$ denotes the adjacency matrix defined by
\begin{equation}
a_{ij}=
\left\{
\begin{array}{ll}
1, & {\mbox{ when}}~ {i~{\mbox{and}}~j~{\mbox{are~linked}}}, \\
0, & {\mbox{otherwise}},
\end{array}
\right.
\end{equation}
and the degree $k_i$ of the vertex $i$ is defined as the number of vertices linked to $i$,
i.e.~$k_i=\sum_j a_{ij}$. In a sparse network, the total number of links
is proportional to $N$.
The second term on the right-hand side of Eq.~(\ref{eq:model})
represents {\em ferromagnetic} coupling to neighboring oscillators on the network ($K>0$),
so the neighboring oscillators favor their phase difference minimized.

When the coupling is weak (small $K$), each oscillator tends to evolve
with its own dynamics described by $d\phi_i/dt\simeq\omega_i$, showing no synchrony.
As the coupling is increased, the oscillators interact with each other and eventually
a macroscopic number of synchronized oscillators can appear if the coupling is
strong enough.
Emergence of macroscopic synchronization can be probed by
the phase order parameter~\cite{ref:Kuramoto}
\begin{equation}
\Delta e^{i\theta}=\frac{1}{N}\sum_{j=1}^N e^{i\phi_j},
\label{eq:Delta-def}
\end{equation}
where $\Delta$ measures the magnitude of the phase
synchrony and $\theta$ denotes the average phase.

Near the synchronization transition ($\Delta\approx 0$), the MF analysis shows that
the order parameter $\Delta$ vanishes with the ordinary MF exponent
$\beta=1/2$, if the degree distribution $\{k_i\}$ is not strongly
heterogeneous~\cite{ref:sync_network_MF,ref:SFN_sync_Hong}. The ER random network has an exponential
degree distribution with weak heterogeneity, thus the ordinary MF exponent should be found.

The FSS is governed by fluctuations. In the system described by Eq.~(\ref{eq:model}),
there could be three different fluctuations. One comes from dynamic (temporal)
fluctuations which persist even in the steady state due to the presence of
running (non-static) oscillators.
Two others are due to {\em quenched} disorder in
link configurations $\{a_{ij}\}$ in a sparse network and possibly in
frequency distributions $\{\omega_{i}\}$ obtained by a random drawing procedure from the
distribution $g(\omega)$. It is difficult to study analytically the dynamic fluctuations even in the globally coupled case (no link disorder) and there still exists a controversy~\cite{ref:gl_noiseless, ref:Daido}. However, it is known that frequency-disorder fluctuations dominate
over dynamic fluctuations at least in the globally coupled case~\cite{ref:Hong_Entrainment, ref:gl_noiseless}.

In this work, we focus on the effects of link-disorder fluctuations separately, so that
we need to remove frequency-disorder fluctuations completely from the system. This can
be achieved by generating the frequencies $\{\omega_{i}\}$ following a deterministic
procedure given by~\cite{ref:gl_noiseless}
\begin{equation}
\frac{i-0.5}{N}=\int_{-\infty}^{\omega_i} g(\omega)d\omega.
\label{eq:noiselessw}
\end{equation}
The generated frequencies are {\em quasi-uniformly spaced} in accordance with the distribution $g(\omega)$ in the population. As this {\em noiseless} frequency set is uniquely
determined, there is no disorder originating from different realizations of frequencies. For the globally coupled model with this
noiseless set, dynamic fluctuations dominate the FSS with the weak
FSS exponent $\bar\nu\simeq 5/4$~\cite{ref:gl_noiseless}, in contrast that
$\bar\nu= 5/2$~\cite{ref:Hong_Entrainment} with disorder in frequency sets by randomly
and independently drawing frequencies from the distribution $g(\omega)$.

The question we raise in this work is what kind of FSS emerges in a sparse network such as
the ER network with the  noiseless frequency set, i.e.~what is the role of disorder in link configurations? Also does the FSS change or not if disorder in frequency sets is added in the ER network? In the following sections, we start with a MF analysis
and report extensive numerical results.

\section{III. Mean-field analysis}
We consider the ER networks as simple examples of uncorrelated random sparse networks.
In the ER random network, any pair of vertices is linked independently with
probability $z/N$.
In the large $N$ limit, the average degree becomes $\langle k \rangle =z$
and the degree distribution is given by the Poisson
distribution~\cite{ref:Erdos, ref:Newman}
\begin{equation}
P(k) =
\frac{\langle k \rangle ^{k}e^{-\langle k \rangle }}{k!} .
\label{eq:Poisson}
\end{equation}
In the $z$-regular random network, all vertices have the same degree $z$ with random
connectivity between vertices, so the degree distribution is given by the
Kronecker $\delta$-function as $P(k)=\delta_{k,z}$.

Taking the annealed MF approximation for connectivity of the network
$a_{ij}\approx k_i k_j /N\langle k\rangle$ (heterogeneous MF theory), one can introduce the global ordering field $H\exp(i\theta)$~\cite{ref:SFN_sync_Hong} as
\begin{equation}
H e^{i\theta}=\frac{1}{N}\sum_{j}\frac{k_j}{\langle k\rangle} e^{i\phi_{j}},
\label{eq:global}
\end{equation}
and then the equation of motion, Eq.~(\ref{eq:model}), can be rewritten as
\begin{equation}
\dot\phi_i = \omega_i - k_iKH \sin(\phi_i - \theta).
\label{eq:mf_eq}
\end{equation}

In the long-time limit where the global ordering field approaches a constant in average,
one can easily solve Eq.~(\ref{eq:mf_eq}) for each  $\phi_i$ with a constant $H$
and $\theta$.
Then, we can set up the self-consistency equation for $H$ through Eq.~(\ref{eq:global}) as
\begin{equation}
H=
 \frac{1}{N}\sum_{j=1}^N \frac{k_j}{\langle
k\rangle} \sqrt{1-\Bigl(\frac{\omega_j}{k_j KH}\Bigr)^2}
\Theta\Bigl(1-\frac{|\omega_j|}{k_j KH}\Bigr), \label{eq:Psi-tilde}
\end{equation}
where $\Theta(x)$ is the Heaviside step function, which takes the value $1$ for
$x\ge 0$ and $0$ otherwise. Each term inside the summation represents the contribution from
an oscillator with $\omega_j$ at vertex $j$ with degree $k_j$.
Note that only {\em entrained} oscillators with  $|\omega_j|<k_j KH$ contribute.

The frequency set $\{\omega_i\}$ fluctuates randomly over the samples
for the noisy frequency distribution, and the degree configuration  $\{k_i\}$ also fluctuates randomly
for the ER random network. Thus, the nontrivial (nonzero $H$) solution of the self-consistent equation
also fluctuates over the samples. With either one of fluctuations, we expect the Gaussian fluctuations
for sufficiently large $N$ from the central limit theorem, and the self-consistency equation
for small $H$ becomes~\cite{ref:SFN_sync_Hong}
\begin{equation}
H=(K/K_c)H-c(KH)^3+d(KH)^{1/2}N^{-1/2}\xi
\label{eq:H}
\end{equation}
with $K_c = [2/\pi g(0)]\langle k \rangle/{\langle k^2 \rangle}$, and constants
$c = -[{\pi g^{''}(0)}/{16}]{\langle k^4 \rangle }/{\langle k \rangle} $ and
$d = \sqrt{[4 g(0)/3]{\langle k^3 \rangle}/{\langle k^2 \rangle}}$.
The term $\xi$ is a Gaussian random variable with zero mean and unit variance,
which represents frequency-disorder fluctuations and/or link-disorder fluctuations.

The scaling solution to Eq.~(\ref{eq:H}) can be easily obtained in the FSS
form~\cite{ref:SFN_sync_Hong, ref:Fisher}
\begin{equation}
H=N^{-1/5}f [ (K-K_c)N^{2/5}],
\label{eq:FSS}
\end{equation}
implying $\beta/\bar\nu=1/5$ and $\bar\nu=5/2$.
We note that this result is the same as that of the globally coupled
{\em noisy} oscillators with frequency-disorder fluctuations~\cite{ref:Hong_Entrainment,ref:gl_noiseless}.

One exceptional case can be constructed when the system is put on the $z$-regular random network
with the noiseless frequency set. In the present MF scheme, there is no fluctuation in the solution of
$H$ and one may expect the same FSS behavior as that of the globally coupled {\em noiseless}
oscillators with $\beta/\bar\nu\approx 2/5$ and $\bar\nu\approx 5/4$, representing dynamic fluctuations only.
However, the MF approximations ignore fluctuations in connectivity disorder (absent for the globally coupled
case), which might give rise to another type of sample-to-sample Gaussian fluctuations
dominant over dynamic fluctuations.

In the next section, we check the validity of the MF analysis via extensive numerical simulations in the ER random and $z$-regular networks with the noiseless and noisy frequency distributions.

\section{IV. Numerical analysis}

We perform extensive numerical simulations for the system governed by Eq.~(\ref{eq:model}).
The ER and the $z$-regular random networks are generated for the average degree $z=\langle k \rangle=6$ up to the
system size $N=12800$. For the {\em noiseless} frequency distribution,
we generate the intrinsic frequencies by the process shown in Eq.~(\ref{eq:noiselessw}) with
$g(\omega)=(2\pi\sigma)^{-1/2}\exp(-\omega^2/2\sigma^2)$ with unit variance $(\sigma^2=1)$.
We used Heun's method~\cite{ref:Heun}, with a discrete time step $\delta t=0.01$, in the
numerical integration of Eq.~(\ref{eq:model}) up to $4\times 10^4$ time steps.
Initial values for  $\{\phi_i(0)\}$ are chosen random, and data are collected
only for the latter half of time steps to avoid any transient behavior~\cite{exp}.
For each network size,  we also average the data over $10^2 \sim 10^3$ different realizations of the network.

\begin{figure}
\includegraphics[width=0.45\textwidth]{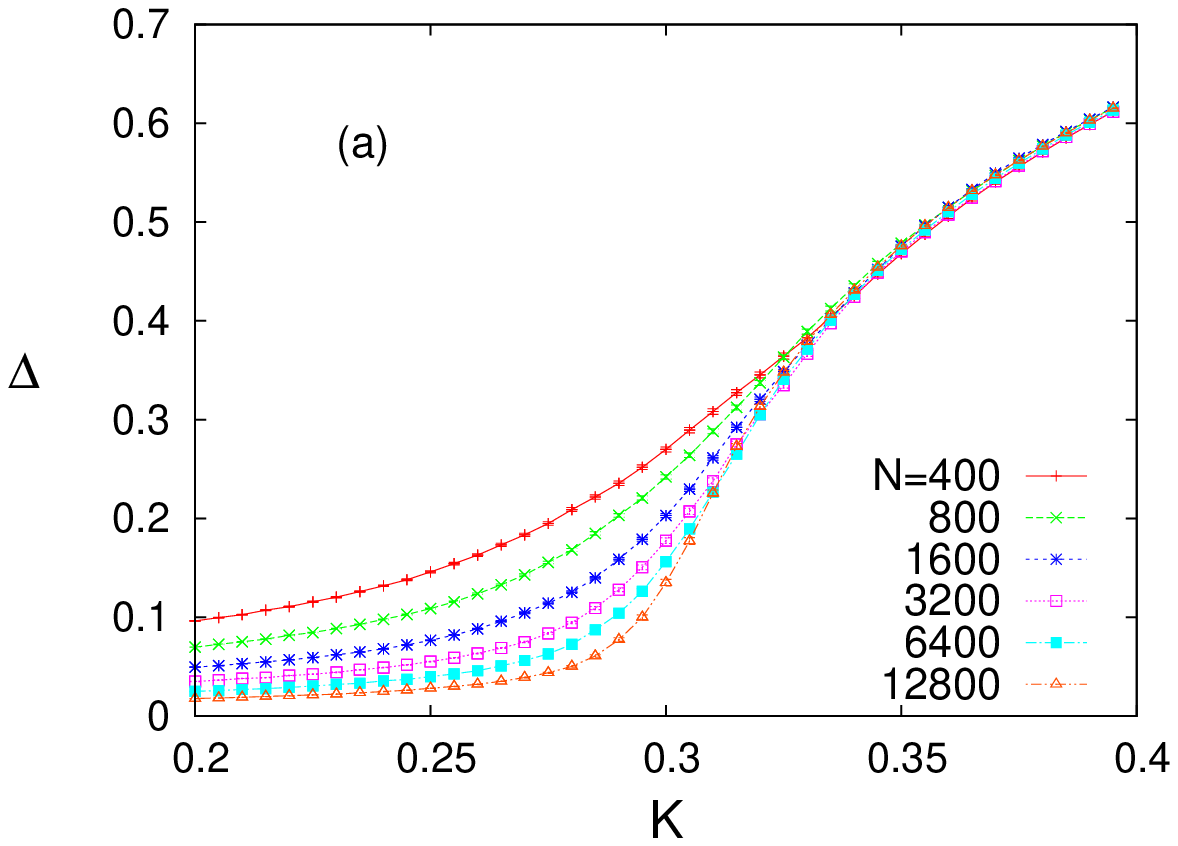}
\includegraphics[width=0.45\textwidth]{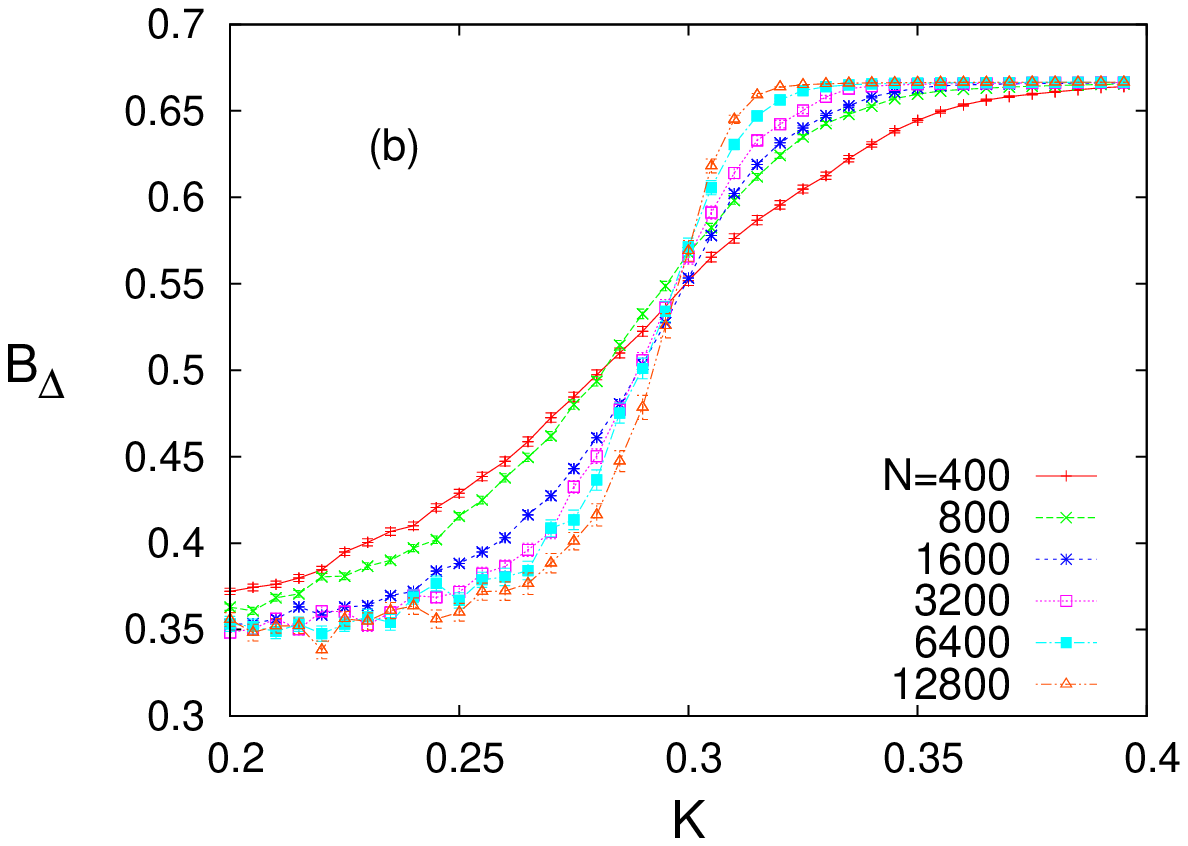}
\caption{(Color online) (a) The order parameter $\Delta$ is
plotted as a function of the coupling strength $K$ for various system size $N$.
(b) Binder cumulant $B_\Delta$ versus $K$ for various system size $N$, where
we estimate  $K_c=0.300(5)$ from the crossing points.}
\label{fig:Delta_FLG}
\end{figure}
First, we consider the ER random network and measure the phase order parameter $\Delta$
as a function of the coupling constant $K$ for various system size $N$,  see Fig.~\ref{fig:Delta_FLG}~(a).
The critical coupling strength $K_c$, beyond which the phase synchronization occurs, can be
accurately located by utilizing the Binder cumulant~\cite{ref:HPT-BC}
\begin{equation}
B_{\Delta}= 1-\Biggr[\frac{\overline{\Delta^4 }}{3\overline{\Delta^2}^2}\Biggr],
\end{equation}
where the overbar represents the time average after sufficient relaxation and
$[\cdots]$ represents the average over different realizations of the networks, respectively.
In the weak coupling regime $K\ll K_c$, the Binder cumulant $B_{\Delta}$
goes to $1/3$ in the large $N$ limit, while it goes to
$2/3$ for the strong coupling regime $(K\gg K_c)$. The Binder cumulant curves
with different $N$ cross each other at the critical coupling strength $K_c$, as seen in
Figure~\ref{fig:Delta_FLG}~(b). We estimate $K_c = 0.300(5)$ by extrapolating
the crossing points in the large $N$ limit. Note that the MF prediction on $K_c$
is about $ 0.23$ in Eq.~(\ref{eq:H}), which is quite far away from the numerical estimate.
It is not surprising to see a considerable shift of the critical point, because
the inherent linking disorder in the quenched networks generates finite correlations in neighboring nodes.
However, it is hoped that the scaling nature near the transition is not affected by finite
correlations, since is is usually universal.

\begin{figure}
\includegraphics[width=0.45\textwidth]{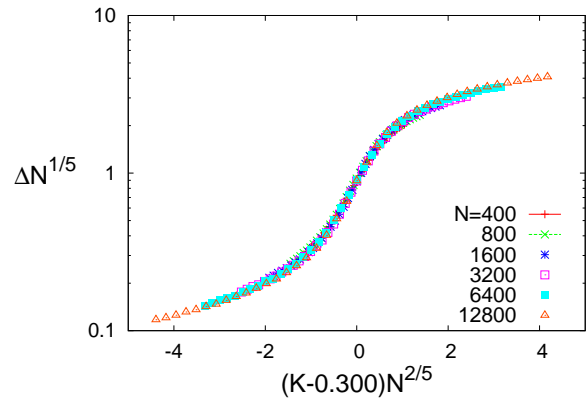}
\caption{ (Color online)
Scaling plot of  $\Delta$ for various system size $N$ in the ER random networks
with the {\em noiseless} frequency distribution.
}
\label{fig:Delta_scaling_FLG}
\end{figure}

\begin{figure}
\includegraphics[width=0.45\textwidth]{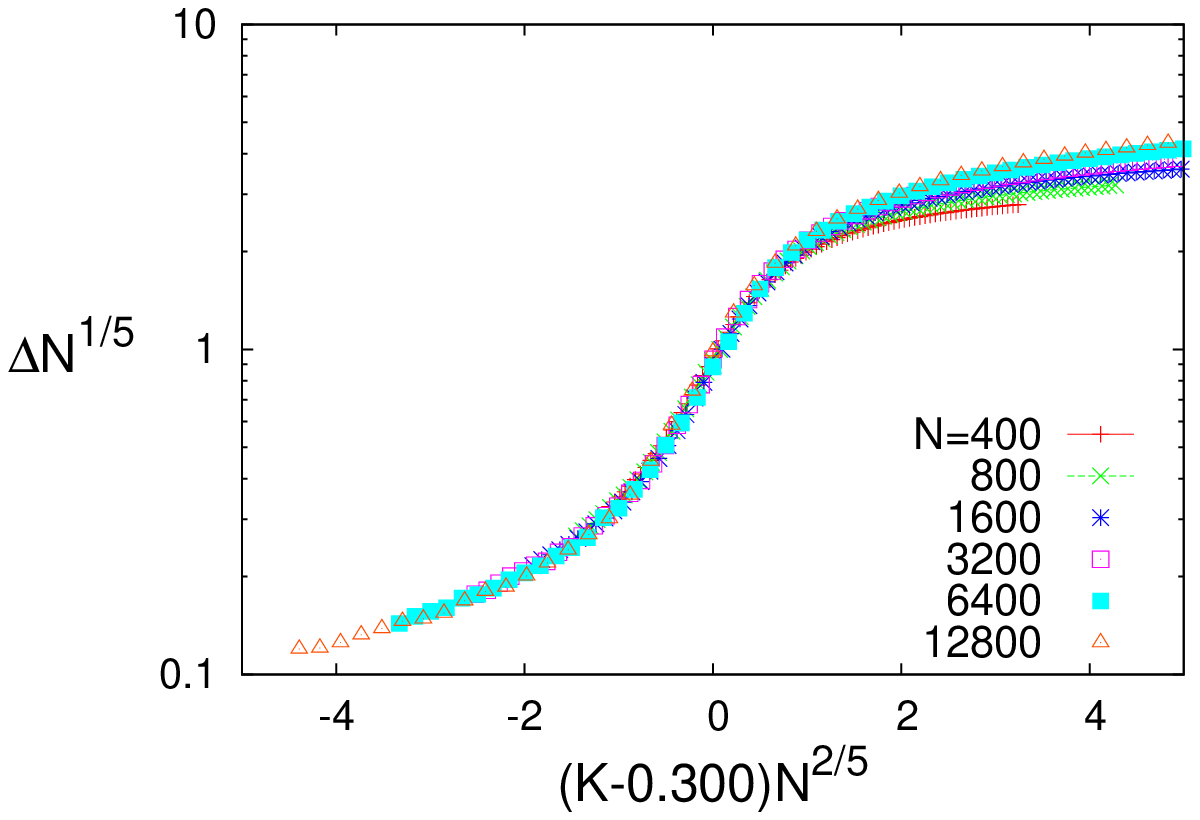}
\caption{ (Color online)
Scaling plot of  $\Delta$ for various system size $N$ in the ER random networks
with the {\em noisy} frequency distribution.
}
\label{fig:Delta_scaling_Gaussian}
\end{figure}

We now check the FSS relation of Eq.~(\ref{eq:FSS}). As one can easily show $\Delta \propto H$ for small $H$ at the MF level~\cite{ref:SFN_sync_Hong}, we expect
\begin{equation}
\Delta N^{1/5}={\tilde f}[(K-K_c)N^{2/5}],
\label{eq:scaling_f}
\end{equation}
where the scaling function ${\tilde f}[x]$ is characterized by
\begin{equation}
{\tilde f}[x]  \sim
\left\{
\begin{array}{ll}
{\mbox {const}}, & x=0, \\
x^{1/2}, & x > 0, \\
(-x)^{-3/4}, & x < 0
\end{array}
\right.
\end{equation}
for the ER random networks with the noiseless frequency distribution.
Figure~\ref{fig:Delta_scaling_FLG} shows the scaling function ${\tilde f}[x]$
by collapsing the order parameter data for various different sizes,
using the exponents $\beta/\bar\nu=1/5$ and $\bar\nu=5/2$, which agrees perfectly with
the MF prediction of the exponent values. Figure~\ref{fig:Delta_scaling_Gaussian}
is the scaling plot for the ER random networks with the noisy frequency distribution.
As expected, the scaling collapse is found with the same values of the exponents
$\beta/\bar\nu=1/5$ and $\bar\nu=5/2$. Moreover, the scaling function ${\tilde f}[x]$
is identical for both the noiseless and noisy distributions with the same critical
coupling strength $K_c$. Differences can be seen only in higher-order finite-size effects.

\begin{figure}
\includegraphics[width=0.45\textwidth]{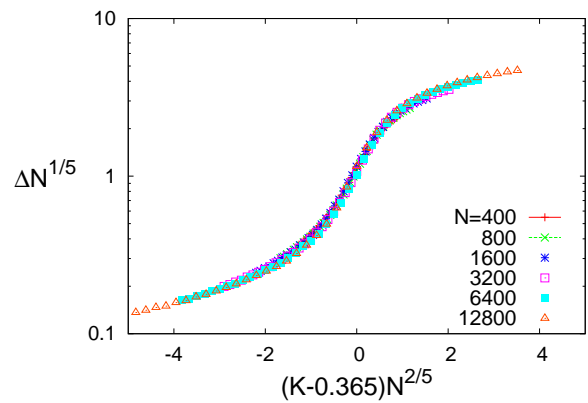}
\caption{ (Color online)
Scaling plot of  $\Delta$ for various system size $N$ in the $z$-regular
random networks with the {\em noiseless} frequency distribution.}
\label{fig:Delta_scaling_regular_FLG}
\end{figure}

We also report the scaling plot for the $z$-regular random network with the
noiseless frequency distribution in Fig.~\ref{fig:Delta_scaling_regular_FLG}.
As the strength of neighboring correlations in quenched networks depends on
the details of networks, the critical point in the $z$-regular network is
located differently from that in the ER random
network, which is estimated as $K_c=0.365(5)$ from the Binder cumulant analysis.
Nevertheless, the corresponding exponent values of $\beta/\bar\nu$ and $\bar\nu$ are identical to those in the
ER random network as shown in Fig.~\ref{fig:Delta_scaling_regular_FLG}, which are
{\em against} the MF prediction discussed in the previous section.
This implies that the random link-disorder fluctuations in the quenched $z$-regular network generate
sample-to-sample Gaussian fluctuations of the same kind in the ER random networks as well
as in the globally coupled network with frequency-disorder fluctuations. However,
the scaling function ${\tilde f}[x]$ by itself is distinct from that in the ER networks.

\section{V. Summary}
In this work, we considered a finite population of coupled random frequency
oscillators on the ER and the $z$-regular random network with link-disorder fluctuations.
In the globally coupled network without any link disorder, the finite-size fluctuations
are determined crucially by the existence of disorder in the intrinsic frequency distribution.
Under the presence of link disorder in random networks, we find that the finite-size
fluctuations are universal with the anomalous FSS exponent $\bar\nu=5/2$,
independent of the details of the degree distribution
(if not strongly heterogeneous) and also independent of the frequency disorder.
We note, however, that the scaling function is not universal, depending on
the details of the network connectivity.

\section{Acknowledgment}
This work was supported by the Mid-career Researcher Program through
the NRF Grant No. 2010-0026627 (H.P.) and the Basic Science Research Program through
No.~2012R1A1A2003678 (H.H.) funded by MEST.


\end{document}